\documentclass[usenatbib]{mnras}

\usepackage{newtxtext,newtxmath}
\usepackage[T1]{fontenc}
\usepackage{ae,aecompl}
\usepackage{booktabs,caption}
\usepackage[flushleft]{threeparttable}

\usepackage{graphicx}	% Including figure files
\usepackage{amsmath}	% Advanced maths commands
\usepackage{amssymb}	% Extra maths symbols
\title[]{A super-Earth and a mini-Neptune around Kepler-59}

\author[Saad-Olivera et al.]{
X.~Saad-Olivera,$^{1}$\thanks{E-mail: ximena@on.br}
C.~F.~Martinez,$^{1}$
A.~Costa de Souza,$^{1}$
F.~Roig$^{1}$
and D.~Nesvorn\'y$^{2,1}$
\\
$^{1}$Observat\'orio Nacional, Rua Gal. Jose Cristino 77, Rio de Janeiro, RJ 20921-400, Brazil\\
$^{2}$Department of Space Studies, Southwest Research Institute, 1050 Walnut Street, Suite 300, Boulder, CO 80302, USA
}

\date{Accepted XXX. Received YYY; in original form ZZZ}

\pubyear{2018}

\begin{document}
\label{firstpage}
\pagerange{\pageref{firstpage}--\pageref{lastpage}}
\maketitle

% Abstract of the paper
\begin{abstract}
We characterize the radii and masses of the star and planets in the Kepler-59 system, as well as their orbital parameters. The star parameters are determined through a standard spectroscopic analysis, resulting in a mass of $1.359\pm 0.155\,M_\odot$ and a radius of $1.367\pm 0.078\,R_\odot$. The planetary radii obtained are $1.5\pm 0.1\,R_\oplus$ for the inner and $2.2\pm 0.1\,R_\oplus$ for the outer planet. 
The orbital parameters and the planetary masses are determined by the inversion of Transit Timing Variations (TTV) signals. For this, we consider two different data sets, one provided by \citet{Holczer2016}, with TTVs only for the planet Kepler-59c, and the other provided by \citet{Rowe2015}, with TTVs signals for both planets. The inversion method is carried out by applying an algorithm of Bayesian inference (\texttt{MultiNest}) combined with an efficient N-body integrator (\texttt{Swift}). For each of the data sets, two possible solutions are found, both having the same probability according to their corresponding Bayesian evidences. All four solutions appear to be indistinguishable within their 2-$\sigma$ uncertainties.  Nevertheless, statistical analyses show that the solutions from \citet{Rowe2015} data  better characterize the data. The first and second solutions identify masses of $5_{-2}^{+4}~M_{\mathrm{\oplus}}$ and $4.6_{-2.0}^{+3.6}~M_{\mathrm{\oplus}}$, and $3.0^{+0.8}_{-0.8}~M_{\mathrm{\oplus}}$ and  $2.6^{+1.9}_{-0.8}~M_{\mathrm{\oplus}}$ for the inner and outer planet, respectively. This points to a system with an inner super-Earth and an outer mini-Neptune. Dynamical studies show the planets have almost co-planar orbits with small eccentricities ($e<0.1$), close but not into the 3:2 mean motion resonance. Stability analysis indicates that this configuration is stable over million years of evolution. 

\end{abstract}


%%%%%%%%%%%%%%%%%%%%%%%%%%%%%%%%%%%%%%%%%%%%%%%%%%

\section{Introduction}
In a single planet system, a transiting planet orbits the star following a Keplerian orbit, periodically blocking the brightness of the star. The presence of more planets in the system turns the orbit to be not perfectly Keplerian, thus the mid-transit times deviate from a linear ephemeris: these deviations are known as Transit Timing Variations (TTVs), and they are particularly sensitive to small  dynamical perturbations between the planets allowing to estimate the planetary masses $M_p$. Moreover, Mean Motion Resonances (MMR) can lead significant perturbations which will turn easier to detect these TTVs \citep{Agol2005, Holman2005}. Knowing the planetary masses and radii from transit observations is fundamental to constraint the planetary densities, which in turn is useful to better understand the planetary formation models.

Here, we present a TTV analysis of planet Kepler-59c that, for the first time, completely characterize the planetary orbits of the two planets in the Kepler-59 system, improving the estimates of their masses. 

The planet Kepler-59c, first identified as KOI-1529.01, was announced from the earliest analyses of the Kepler data (quarters Q1-Q2; \citealp{Borucki2011}). The analysis of data in quarters Q1-Q6 showed a second planet candidate, identified as KOI-1529.02. \citep{Batalha2013}. Using Kepler data from quarters Q1-Q8, \citet{Steffen2013} revealed the presence of anti-correlated TTVs by dynamical modelling coupled with Monte Carlo analysis, confirming the planetary nature of the two candidates and that they orbit the same star. 
Using the stellar parameters of Kepler-59 ($M_\star=1.04~M_\odot$ and $R_\star=0.94~R_\odot$) and the ratios of planet-to-stellar radius for the two planets ($R_b/R_\star=0.01065\pm5.3x10^{-4}$ and $R_c/R_\star=0.01923\pm7.6x10^{-4}$) given in \citet{Batalha2013}, \citet{Steffen2013} computes the radius of planet b as $R_b=1.09\pm0.05~R_{\mathrm{\oplus}}$ and of planet c as $R_c=1.97\pm0.08~R_{\mathrm{\oplus}}$. Taking into account the uncertainties on this parameter, it is possible to classify Kepler-59b as an Earth planet, and Kepler-59c as either a super-Earth or a mini-Neptune planet. Nevertheless more study is needed in order to analyse the internal structure of these small planets to give an affirmation. The last two kind of objects where found to be galore in the Kepler observations, and demonstrate to be common in our Galaxy \citep{Batalha2013}, occurring with orbital periods between 5 and 50 days \citep{Petigura2013, Silburt2015}.

The TTVs analysis together with the dynamical study of the Kepler-59 system enable \citet{Steffen2013} to set upper limits to the planetary masses as to guarantee the stability of the system. Indeed, for the system to be stable the mass of Kepler-59b has to be $M_b<2.05~M_{Jup}$, and the mass of Kepler-59c has to be $M_c<1.37~M_{Jup}$. 

Our goal in this paper is to improve the values of the masses of both planets in the Kepler-59 system, not by defining upper limits but by accurately constraining them. This is done first using the TTV data of Kepler-59c calculated by \citet{Holczer2016} from Kepler quarters Q1-Q17, and then using the TTV data of Kepler-59b and c calculated by \citet{Rowe2015} from quarters Q1-Q12. 

In Section~\protect{\ref{star}}, we estimate the stellar parameters and derive the planetary radius from the light curve analysis. In Section~\protect{\ref{methods}}, we explain the methodology used to fit the mid-transit times of the planets. The results are shown in Section~\protect{\ref{results}}, together with a dynamical study of the obtained solution.
Discussions and conclusion are given in Section~\protect{\ref{discu}}.

%%%%%%%%%%%%%%%%%%%%%%%%%%%%%%%%%%%%%%%%%%%%%%%%%%%%%%%%%%%%%%%%%%%%%%%%
\section{Stellar and light curve parameters}
\label{star}
\subsection{Spectroscopic Stellar Parameters}
The high-resolution spectrum of Kepler-59 analyzed in this work was obtained with the High Resolution Echelle Spectrometer \citep[HIRES,][]{vogt1994} at the Keck telescope as part of the large observational campaign named California-Kepler Survey \citep[CKS-][]{petigura2017,johnson2017}, which is focused on target stars identified as Kepler Objects of Interest (KOIs). 
The reduced spectrum is publicly available and was obtained from the Keck Observatory Archive {\tt https://california-planet-search.github.io/cks-website/}. We estimated the spectral signal-to-noise ratio (S/N) by measuring the rms-flux fluctuation in selected continuum windows with typical values of S/N between $\sim 35$-40. 
We conducted a classical spectroscopic analysis, based on equivalent widths (EW) measurements of a selected set of Fe I and Fe II lines to derive the stellar parameters: effective temperature, $T_{\rm eff}$, surface gravity, $\log g$, and micro-turbulent velocity, $\xi$, as well as the metallicity [Fe/H] of the Kepler-59 star.
The parameter calculations were done under the assumption of local thermodynamic equilibrium (LTE) using 1D model atmospheres from the {\tt Kurucz ATLAS9 ODFNEW} grid \citep{castelli2004} and the revised version (2002) of the abundance analysis code {\tt MOOG} \citep{sneden1973} to compute the iron abundances.
The stellar parameters were obtained by using the LTE approach and iterating until: (i) the Fe I abundance, A(Fe I), shows no correlation with the excitation potential, EP, of the Fe I line transitions (excitation equilibrium), (ii) the value of 
A(Fe I) exhibits no dependence with the reduced equivalent widths $\log(\mathrm{EW}/\lambda$), and (iii) the mean abundances obtained by Fe I and Fe II lines reach similar values (ionization equilibrium). These three conditions define the stellar $T_{\rm eff}$, $\log g$, $\xi$ and [Fe/H]. 
We adopted the line list consisting of 158 Fe I and 18 Fe II isolated and unblended lines presented in \cite{ghezzi2018}. The log $gf$ values for the Fe I and II lines were obtained by these authors performing an inverted solar analysis using a {\tt Kurucz ATLAS9 ODFNEW} model atmosphere for the Sun ($T_{\rm eff} = 5777$ K, $\log g = 4.44$, [Fe/H] = 0.00 and $\xi = 1.00$ km~s$^{-1}$), and adopting a solar iron abundance of 7.50 \citep{asplund2009}. 
The line EWs of Kepler-59 were measured using the {\tt IRAF}\footnote{{\tt IRAF} is distributed by the National Optical Astronomy Observatory, which is operated by the Association of Universities for Research in Astronomy, Inc., under cooperative agreement with the National Science Foundation.} $splot$ task.
The error in the derived effective temperature was calculated by changing this parameter until the coefficient correlation between the A(Fe I) and EP achieves the value of the slope uncertainty from the converged solution. The error for $\xi$ was calculated in a similar way, but using the coefficient correlation between the A(Fe I) and $\log(\mathrm{EW}/\lambda$) instead. To estimate the uncertainties in $\log g$, we varied this parameter until the Fe I and Fe II mean abundances differed exactly by one standard deviation of the mean of A(Fe I). Finally, the [Fe/H] error was obtained by adding in quadrature the standard deviation of the mean A(Fe I) and the variations caused by the uncertainties in $T_{\rm eff}$ and $\log g$.

\subsection{Stellar Radius and Mass}
The stellar radius was obtained by using the Stefan-Boltzmann Law depending on the effective temperature (already spectroscopically derived) and the stellar luminosity. To calculate the latter, we first obtained the absolute magnitude by considering: (i) the 2MASS K-band, (ii) the $A_{k}$ extinction, after transforming the reddening $E(\mathrm{B}-\mathrm{V})$ derived from the 3D dust map of \cite{green2018} using the relations from \cite{bilir2008}, and (iii) the distance modulus, with the distance being estimated by \cite{bailer2018} after performing a Bayesian approach assuming geometric priors with the Gaia DR2 parallaxes \citep{Gaia2018,Gaia2016}, and taking into account the systematic parallax offsets determined from Gaia's observations of quasars \citep{lindegren2018,Zinn2017}. The bolometric magnitude was then obtained by adding the bolometric correction to the absolute magnitude, derived by using the {\tt isoclassify} package \citep{huber2017}, which interpolates over the MIST tables \citep{choi2016}. Once we had the stellar luminosity, $L_{\star}$, we combined it with our measured $T_{\rm eff}$ to finally derive the stellar radius, $R_{\star}$.
We also estimated the stellar mass ($M_{\star}$) using our spectroscopically well-constrained $\log g$ obtained via an equivalent width spectroscopic technique, and our derived $R_{\star}$ leveraging the precision achieved by using the {\tt Gaia} parallax.

\begin{table}
    \centering  
    \begin{threeparttable}
	  \caption{Stellar properties derived from spectroscopic analysis.}
	   \label{tab:star59}
             \begin{tabular}{llc}
              \toprule
             Parameter      & Value  & Reference  \\ [0.5ex] 
             \midrule
             $m_{Ks}$ & $ 12.928 $ & SO18 \\
                       & $ 12.928 $ & T18 \\
             \vspace{0.1cm}
             $A_{Ks}$ & $0.00794 $ & SO18 \\
             \vspace{0.1cm}
             $BC$ & $ 1.282 $  & \\
             \vspace{0.1cm}
             $T_{eff}$(K) & $ 6050^{+100}_{-100} $  & SO18 \\
                          & $ 6296^{+169}_{-207} $  & T18 \\
                          & $ 5884^{+118}_{-118} $ & Be18 \\
                          & $ 6074 $                & M16 \\
                          & $ 6074 $                & S13 \\
                          & $ 6074 $                & Ba13 \\
             \vspace{0.1cm}
             $log(g)$ & $ 4.3^{+0.18}_{-0.18} $     & SO18 \\
                      & $ 4.46^{+0.052}_{-0.221} $  & T18 \\                
                      & $ 4.350^{+0.091}_{-0.190} $ & M16 \\                
             \vspace{0.1cm}
             $r$ [pc] & $ 1163.221^{+22.136}_{-21.339} $      & SO18 \\ 
                      & $ 1162.742^{+21.664}_{-20.943}$       & B18 \\
                      & $ 1203.573^{23.094}_{23.094}$         & G18 \\
                      & $ 1021.000^{+308.374}_{-127.418} $    & M16 \\
             \vspace{0.1cm}
             $[Fe/H]$[dex] & $ -0.13^{+0.1}_{-0.1} $        & SO18 \\
                           & $ -0.26^{+0.25}_{-0.35} $      & T18  \\
                           & $ 0.0200^{+0.1410}_{-0.1690} $ & M16  \\
             \vspace{0.1cm}                   
             ${R_{\star}}$($R_\odot$) & $ 1.367^{+0.078}_{-0.078} $ & SO18 \\
                                      & $ 1.006^{+0.312}_{-0.111} $ & T18  \\
                                      & $ 1.373^{+0.064}_{-0.061} $ & Be18 \\
                                      & $ 1.170^{+0.352}_{-0.139} $ & M16 \\
                                      & $ 0.94 $                    & S13 \\
                                      & $ 0.94 $                    & Ba13 \\
             \vspace{0.1cm}    
             $M_{\star}$($M_\odot$) & $ 1.359^{+0.155}_{-0.155}$  & SO18 \\
                                    & $ 1.066^{+0.143}_{-0.143} $ & T18 \\
                                    & $ 1.120^{+0.106}_{-0.077} $ & M16 \\
                                    & $ 1.04 $                    & S13 \\
                                    & $ 1.04 $                    & Ba13 \\
             \bottomrule
    \end{tabular}
             \begin{tablenotes}
             \small 
              \item The reference are: Ba13 for \citep{Batalha2013}, S13 for \citep{Steffen2013}, M16 for \citep{Morton2016}, Be18 for \citep{Berger2018}, G18 for \citep{Gaia2018}, T28 for \citep{Thompson2018} and SO18 for this Work.
             \end{tablenotes}
  \end{threeparttable}
\end{table}
To compute the stellar radius and mass uncertainties, we performed an error propagation taking into account the error contributions of all the parameters involved. See Table~\protect{\ref{tab:star59}} for the derived parameters of Kepler-59 and their corresponding uncertainties. 

%%%%%%%%%%%%%%%%%%%%%%%%%%%%%%%%%%%%%%%%%%%%%%%%%%%%%%%%%%%%%%%%%%%%%%%%
\subsection{Planetary Radius}
\label{radius}
The planetary radius is not directly observed; rather, the transit depth, $\Delta F_p$, is the observable from the light curve which is then related to the planet size. The last Data Release (DR15) from Kepler data showed transit depth values for Kepler-59 b and c of $\Delta F_b = 101.9^{+8.4}_{-8.4}$~ppm and $\Delta F_c =  213^{+12.2}_{-12.2}$~ppm  \citep{Thompson2018}. Hence, we calculate the planetary radius $R_p$ as those values obtained from considering that $\Delta F_p = (R_p/R_\star)^2$ and the uncertainties associated were calculated from error propagation. For planet b we obtain a radii of $R_b=1.5^{+0.1}_{-0.1}$~$R_{\mathrm{\oplus}}$ and for planet c a radii of  $R_c=2.2^{+0.1}_{-0.1}$ ~$R_{\mathrm{\oplus}}$ \footnote{ The planetary radius determination depends on the assumed stellar radius. Hence, different values for the stellar radius can give larger or smaller planetary radii.}. 

Taking into account this values it is almost clear that the planetary nature of this objects corresponds to a super-Earth and a mini-Neptune planets, as seen in previous work \citep{Steffen2013}. Following we derived the mass for both planets in order to corroborate this affirmation.

%%%%%%%%%%%%%%%%%%%%%%%%%%%%%%%%%%%%%%%%%%%%%%%%%%%%%%%%%%%%%%%%%%%%%%%%
\section{Inversion method}
\label{methods}
The inversion method applied to the Kepler-59 system uses the algorithm known as {\tt MultiNest} \citep{Feroz2009,Feroz2013}, which relies on the Bayes rule to infer the parameters of the planetary system that better fit the observations. This algorithm also calculates the evidence term of the Bayes rule, through a Nested Sampling routine, allowing us to select between different model parameters that fit the data.
The planetary systems are simulated using the N-body dynamical integrator {\tt Swift} \citep{Levison1994}, that accounts for the gravitational interactions between all the bodies in the system. The code is adapted to provide the mid-transit times of the planets \citep{Nesvorny2013}, and it generates transits for each system using the model from \citet{Mandel2002}. We choose to apply the inversion method to the mid-transit times, instead of the TTVs, because the Transit Timing Variations are nothing more than the mid-transit times minus a linear ephemeris that fit the mid-transit times. In this way the step to calculate the TTVs is left aside in the inversion method in order to not carry its associated uncertainties. Our combined code {\tt MultiNest}+{\tt Swift} is written in {\tt Fortran 90} and can be parallelized with {\tt Open MPI}. 

For the two data sets analyzed in this work, i.e. mid-transit times of only Kepler-59c \citep{Holczer2016}, and mid-transit times of both Kepler-59b and c \citep{Rowe2015}, we assume planetary systems where the host star and only two planets are present. Table~\protect{\ref{tab:priors59}} shows the priors chosen for the 13 parameters use to perform the inversion analysis in both cases. The parameters are $M_b/M_\star$, $M_c/M_\star$, $P_b$, $P_c$, $e_b$, $e_c$, $b_b$, $b_c$ $\varpi_b$, $\varpi_c$, $\lambda_b$ or $\delta t_b$, $\delta t_c$, and $\Omega_b-\Omega_c$.
The use of $\lambda_b$ or $\delta t_b$ depends on whether the mid-transits times of planet b are available or not.
The stellar parameters are input parameters of the algorithm, and are kept fixed during the inversion. The priors of the planetary periods are taken from the work of \citet{Steffen2013}, which constraints period values of $P_b=11.86$ days and $P_c=17.9$ days. The impact parameters, $b$, of both transiting planets are obtained from the inversion method, and their priors range between 0 (central transit) and 1 (graze transit). The ration of planetary-to-stellar masses, $M/M_{\star}$, can take prior values ranging from 0 to 7.28~$M_{J}$. The angular position of the orbits determined by $\varpi$, $\lambda$ and $\Omega$ are describe in the reference system where $\Omega_c=270^{\circ}$, as in \citet{Nesvorny2012}, and their priors range between 0 and $360^\circ$. We choose all priors to follow uniform distributions.

 For a more consistent work and in order to not impose strictly priors on the planetary periods we also use wider priors for both planets in both cases analysed here (eg, $P_b=$[1-60] and $P_c=$[1-80]). The results obtained with this priors are comparable with the ones obtained with more reduced priors within 1-$\sigma$ uncertainty. This show that for this system it is possible to imposed a strong prior in the planetary period of b around the solution found from transit light curve directly.

In the next session we only show the results for the priors values list in Section \ref{tab:priors59}. 
\begin{table}
	\caption{Prior distributions of our model, for the two data sets analyzed here. The intervals represent the minimum and maximum values of the uniform distributions for each planet (the sub-index $p$ refers to any of the planets).}
	\label{tab:priors59}
    \centering
             \begin{tabular}{lcc}
              \toprule
               &  Kepler-59c TTVs  & Kepler-59b and c TTVs  \\ [0.5ex]
             \midrule
             $M_p/M_\star$ & $[0.0,\,0.005]$ & [0.0,\,0.005] \\
             \vspace{0.1cm}
             $P_b$ (d) & $[11.7,\,11.9]$ & $[11.7,\,11.9]$\\
             \vspace{0.1cm}
             $P_c$ (d) & $[17.7,\,18.0]$  & $[17.7,\,18.0]$ \\
             \vspace{0.1cm}
             $e_p$ & $[0,\,0.5]$ & $[0,\,0.5]$ \\                
             \vspace{0.1cm}
             $b_p$ & $[0,\,1]$  & $[0,\,1]$ \\
             \vspace{0.1cm}             
             ${\varpi}_p$~($^\circ$) & $[0,\,360]$ & $[0,\,360]$ \\
             \vspace{0.1cm}    
             $\lambda_b$~($^\circ$) - $\delta t_b$~(d) & $[0,\,360]$ & $[1,\,1.2]$ \\
             \vspace{0.1cm} 
             $\delta t_c$~(d) & $[0,\,0.05]$ & $[0,\,0.02]$ \\ 
             \vspace{0.1cm} 
             $\Omega_b-\Omega_c$~($^\circ$) & $[0,\,360]$ & $[0,\,360]$ \\
             \bottomrule
             \end{tabular}                
\end{table}

%%%%%%%%%%%%%%%%%%%%%%%%%%%%%%%%%%%%%%%%%%%%%%%%%%%%%%%%%%%%%%%%%%%%%%%%%%%%%%%%%%%%%%%%%%%%%%%%%%%%%%%%%%%%%%%%%%%%%%%%%%%%
\section{Results}
\label{results}

\subsection{Kepler-59c TTVs }
\label{results1}
We first apply our inversion method to the Kepler-59c TTV data from \citet{Holczer2016}. Holczer present the full list of 72 transits of this planet, and point out 9 transits as being outliers. Using the full data set,
we run MultiNest setting efficiency parameter efr = 0.1, convergence tolerance parameter tol = 1.0, multi-modal parameter mmode = True, random seed control parameter seed = -1, and the constant efficiency mode ceff = False. The number of live samples $N_{live}$ is 4000. We find a bimodal posterior, i.e. two modes in the solution.
 \begin{table*}
	\caption{Parameters estimated by solution 1 ($S^H_1$) and solution 2 ($S^H_1$) from \citet{Holczer2016}.
    The orbital parameters are the osculating astrocentric elements at epoch BJD~2\,455\,052. The upper block are the mean values and the error bars reported from the dynamical fit parameters, at the standard 68.34\% confidence level. The lower block reports the derived parameters.}
	\label{tab:param59H}
    \centering  
             \begin{tabular}{lcccc}
             \toprule
 &\multicolumn{2}{|c|}{$S^H_1$}  & \multicolumn{2}{|c|}{$S^H_2$}  \\ [0.5ex] 
           & \bf{Kepler-59b}  &  \bf{Kepler-59c} & \bf{Kepler-59b}  &  \bf{Kepler-59c} \\ [0.5ex] 
               \midrule
              Dynamical fit            &  &     \\ [0.5ex] 
                $M_p/M_\star\,(\times 10^{-3})$ & $0.15^{+0.06}_{-0.04}$ & $0.4^{+0.3}_{-0.2}$ & $0.07^{+0.01}_{-0.01}$ & $0.05^{+0.03}_{-0.02}$ \\
                \vspace{0.1cm}
                $P_p$~(d) &$11.879^{+0.009}_{-0.007}$ & $17.970^{+0.001}_{-0.002}$& $11.869^{+0.003}_{-0.002}$ & $17.969^{+0.002}_{-0.002}$ \\
                \vspace{0.1cm}
                $e_p$ & $0.03^{+0.03}_{-0.02}$ & $0.02^{+0.02}_{-0.01}$ & $0.07^{+0.02}_{-0.02}$ & $0.04^{+0.02}_{-0.02}$  \\
                \vspace{0.1cm}
                $b_p$  & $0.4^{+0.3}_{-0.2}$ & $0.5^{+0.3}_{-0.3}$ & $0.4^{+0.3}_{-0.2}$ & $0.5^{+0.3}_{-0.3}$ \\
                \vspace{0.1cm}
                $\varpi_p$~($^\circ$) & $209^{+109}_{-172}$ &  $74^{+65}_{-150}$ & $296^{+115}_{-73}$ &  $65^{+70}_{-124}$\\
                \vspace{0.1cm} 
                $\lambda_p$~($^\circ$) & $113^{+177}_{-47}$ & --  & $155^{+163}_{-76}$ & -- \\
                \vspace{0.1cm} 
                $\Omega_p$~($^\circ$) & $272^{+117}_{-121}$ & $270$  & $271^{+117}_{-115}$  & $270$  \\
                \vspace{0.1cm} 
                $\delta t$~(d) & --  & $0.0206^{+0.007}_{-0.007}$  & --  & $0.027^{+0.007}_{-0.007}$ \\    
               \midrule        
             Derived parameters    &  &     \\ [0.5ex] 
                \vspace{0.1cm}                
                $M_p$~($M_{\earth}$) & $ 7^{+3}_{-2} $ & $20^{+13}_{-12}$ & $3.5^{+0.6}_{-0.7} $ & $2^{+1}_{-1} $  \\
                \vspace{0.1cm}                               
                $a_p$~(au)  & $0.112^{+0.002}_{-0.002}$ & $0.148^{+0.002}_{-0.002}$  & $0.112^{+0.002}_{-0.002} $ & $0.148^{+0.002}_{-0.002} $ \\     
                \vspace{0.1cm} 
                $i_p$~($^\circ$)  & $88.6^{+1.0}_{-0.8}$ & $88.7^{+0.7}_{-0.7}$ & $88.9^{+0.9}_{-0.8}$ & $89.1^{+0.7}_{-0.7}$ \\   
                \vspace{0.1cm}                                 
                $I_p$~($^\circ$)  & $1.3^{+1.0}_{-0.8}$ & $1.2^{+0.7}_{-0.7}$  & $1.01^{+0.97}_{-0.83}$ & $0.85^{+0.74}_{-0.73}$ \\   
                \vspace{0.1cm}                                 
                $I_{\mathrm{mut}}$~($^\circ$)  & -- & $0.12^{+1.27}_{-0.12}$ & -- & $0.15^{+1.21}_{-0.15}$ \\   
                \vspace{0.1cm}                                 
                $R_p$~($R_{\mathrm{\oplus}}$)  & $1.5^{+0.1}_{-0.1}$ &  $2.2^{+0.1}_{-0.1}$  & $1.5^{+0.1}_{-0.1}$ &  $2.2^{+0.1}_{-0.1}$ \\
                \vspace{0.1cm}
                $\rho_p$~(g\,cm$^{-3}$)  &  $11^{+4}_{-5}$ & $11^{+7}_{-9}$  &  $5.6^{+1.0}_{-1.7}$ & $1.1^{+0.7}_{-0.9}$ \\
             \bottomrule               
               \end{tabular}                
\end{table*}
\begin{figure*}
    \centering
	\includegraphics[width=6cm,trim={3.2cm 0 3.5cm 0cm}]{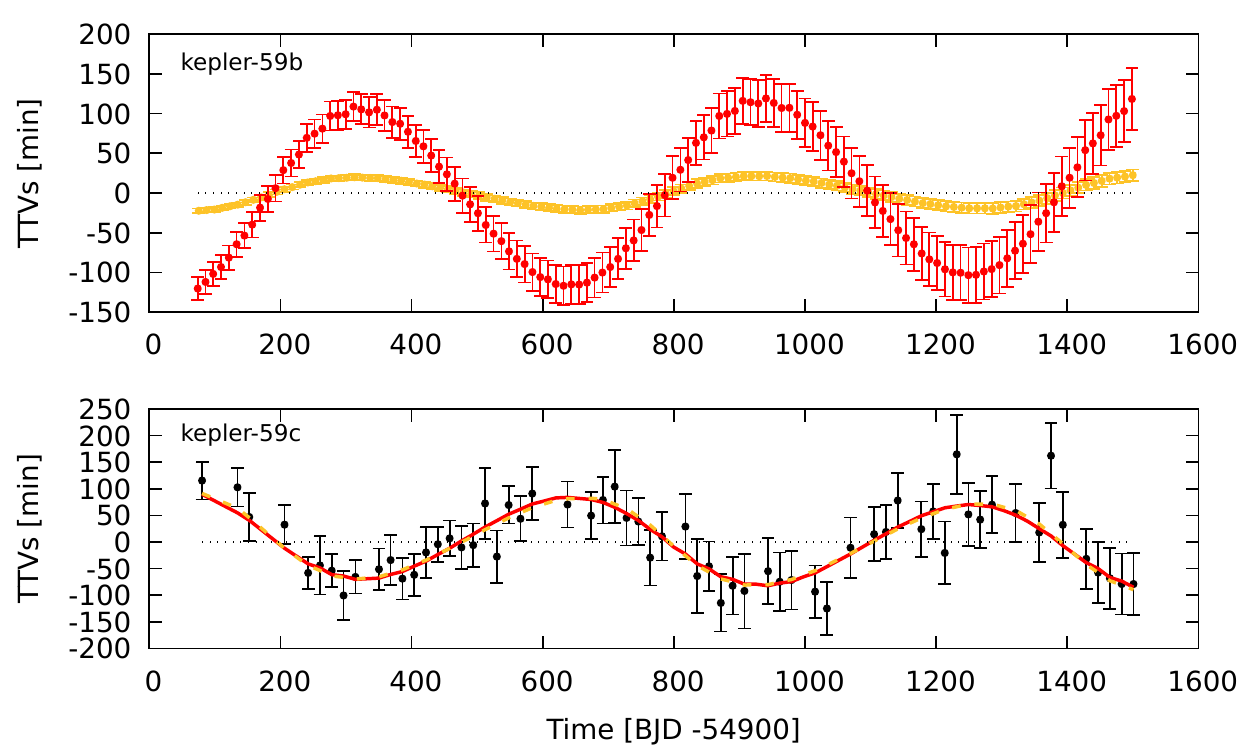}
	\caption{TTVs of planets Kepler-59 b and c. The bottom panel shows the TTVs reported by \protect\citet{Holczer2016} as black dots with error bars and the best fit of solution $S^H_1$ and $S^H_2$ (see Table \ref{tab:param59H}) are the red and orange lines, respectively. The top panel shows the TTVs of planet b generated by the solution $S^H_1$ (red dots) and TTVs of planet b generated by the solution $S^H_2$ (orange dots).}
    \label{fig:O-C59}
\end{figure*}
The first mode represents the first solution, hereafter $S^H_1$, obtained with an evidence of $\ln(Z)_{S^H_1}=128.10$. The parameters provided by {\tt MultiNest} are showed in the top of Table~\ref{tab:param59H}, and are referred as the dynamical fit parameters corresponding to the 13 parameters in Table~\protect{\ref{tab:priors59}}. At the bottom of Table~\ref{tab:param59H}, we present the derived parameters for both planets: the planetary mass $M_p$ in Earth mass units; the semi-major axis $a_p$ in au; the orbital inclination with respect to the sky plane at transit, $i_p$; the inclination of the orbit with respect to the transit plane, $I_p$ (i.e. the plane where $b=0$), and the mutual orbital inclination of the orbits, $I_{\mathrm{mut}}$. The fit points out to an inner planet in an orbit with a period of $P_b=11.879_{-0.007}^{+0.009}$ days and a mass of $M_b=7.11_{-2.1}^{+3.0}~M_{\mathrm{\oplus}}$, while the outer planet has an orbit with $P_c=17.970_{-0.002}^{+0.001}$ days and a mass of $M_c=20.9_{-12.5}^{+13.6}~M_{\mathrm{\oplus}}$. The eccentricities and mutual inclination of the planets indicate that they are in nearly circular and nearly co-planar orbits. The corresponding fit of this solution to Holczer's TTVs is shown in the bottom panel of Figure~\protect{\ref{fig:O-C59}}.

The second mode or second solution, hereafter $S^H_2$, has an evidence value of $\ln(Z)_{S^H_2}=128.56$. By comparing the values of $\ln(Z)$ between the two solutions, we are not able to prefer one solution over the other. Therefore, it is important to  highlight the differences between both solutions. The dynamical fit and derived parameters of solution $S^H_2$ are shown in Table~\protect{\ref{tab:param59H}}. We can see that the planetary periods are indistinguishable from those of solution $S^H_1$ when considering their 1-$\sigma$ confidence levels. This second solution shows again that the planets are close to a MMR, but the behavior of the planetary masses is opposite to that of solution $S^H_1$. The estimated masses are $M_b=3.5^{+0.6}_{-0.7}~M_{\mathrm{\oplus}}$ and  $M_c=2.1^{+1.4}_{-1.2}~M_{\mathrm{\oplus}}$, hence Kepler-59b appears to be more massive than Kepler-59c. The eccentricities do not show significant differences with respect to the first solution. The same happens to the inclinations of the system. The fit of this second solution to the Holczer's TTVs is shown in the bottom panel of Figure~\protect{\ref{fig:O-C59}}.  

Regarding the periods of both solutions, we can see that the perturbation over planet c is produced by a planet in an $11.879^{+0.009}_{-0.007}$ days orbit, or in an $11.869^{+0.003}_{-0.002}$. If we compare this values to the period of planet b obtained from its transits \citep{Steffen2013}, we get $P_b=11.8681^{0.0003}_{0.0003}$ days, and we verify that the values are indistinguishable within the 1-$\sigma$ uncertainties. This would mean that the inner perturber corresponds to an actual planet in that position.

From both solutions we can generate the mid-transit times of planet b and the TTVs that such planet would be displaying. At the top panel of Figure~\protect{\ref{fig:O-C59}} in red, we show the TTVs of planet b that are expected from $S^H_1$.  This solution shows an inner planet displaying a large TTV amplitude ($\sim100$ minutes). In the same Figure the TTVs of planet b expected from solution $S^H_2$ is shown. This second solution shows a TTV signal with lower amplitude when compared to that of the first solution. This is a consequence of different planetary masses: the lower the mass of the planet, the lower the perturbation on the companion, and vice-versa. 

Finally, we also apply the inversion method to the data, but this time taking out the 9 outliers pointed out by Holczer. Once again, we find two solutions. We do not show them because they are comparable to the solutions $S_1^H$ and $S_2^H$ within the 1-$\sigma$ uncertainties.

\subsubsection{TTVs periodicity }
Defining the planetary period ratio of both solutions to be $P_c/P_b$, we obtain values of $\sim 1.51$, which point to orbits barely outside of the 3:2 mean motion resonance (MMR). \citet{Lith2012} showed that systems close to first order MMR ($\equiv j:j-1$) exhibit sinusoidal TTVs signals, and that the libration period of the signal, referred to as ``super-period'', is inversely proportional to the distance to the resonance. This is expressed as
\begin{equation}
    P^{j}\equiv \displaystyle\left|\frac{j}{P_c}-\frac{j-1}{P_b}\right|^{-1}
\label{eq:superperiod}    
\end{equation}
The super-periods of both solutions found here are
\begin{equation}
    P_{S^H_1}=704.5~ \mathrm{d} \quad\ \text{ and } \quad\ P_{S^H_2}=644.3~ \mathrm{d}
\end{equation}

\begin{figure}
    \centering
	\includegraphics[width=8.0cm]{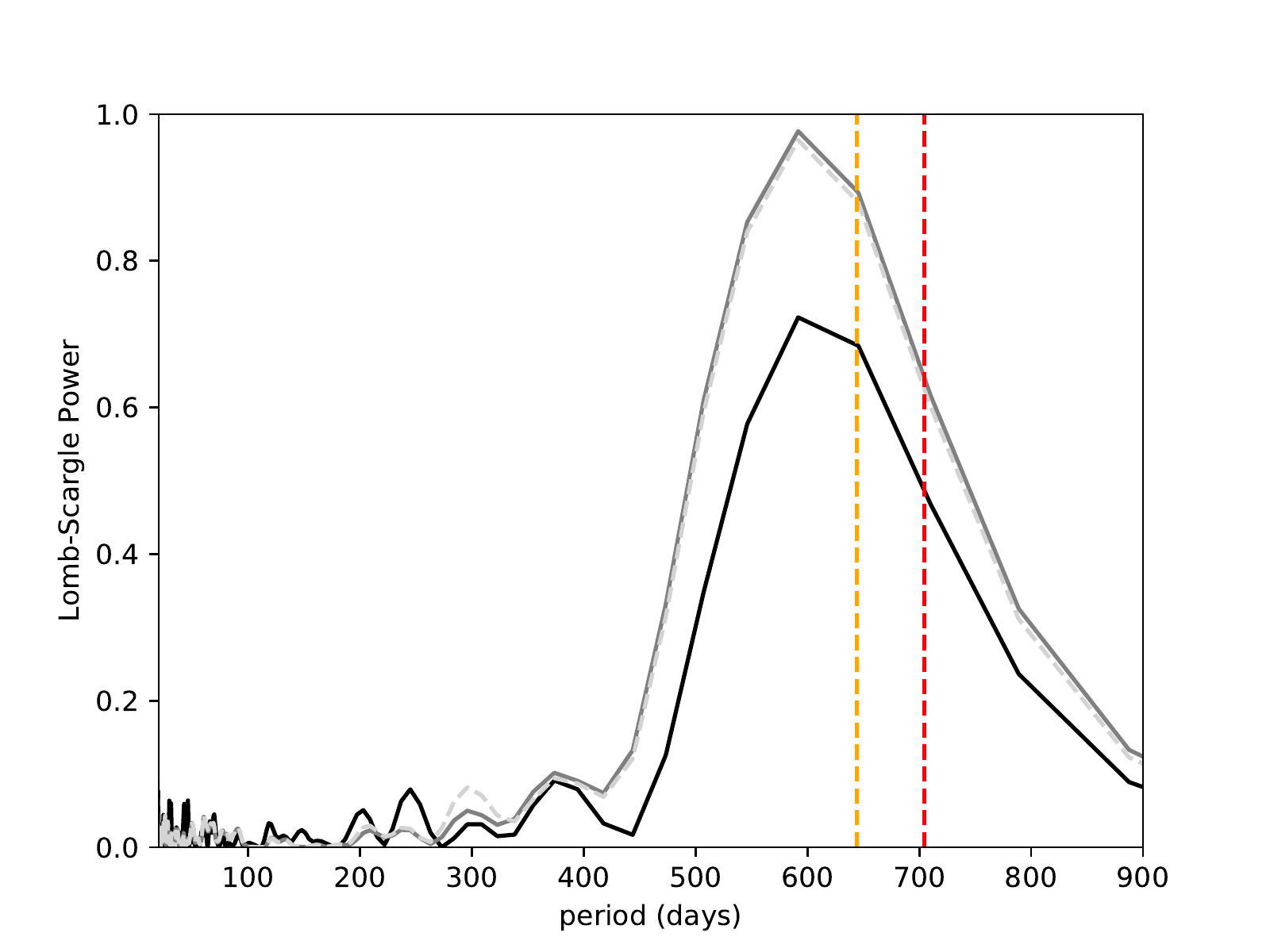}
	\caption {Power spectrum period of Holczer's data for Kepler-59c (black full line), for the solution $S^H_1$ (grey full line), and for the solution $S^H_2$ (light grey dashed line). The vertical dashed lines are the analytic super-period estimates for both solutions (red line for $S^H_1$ and orange line for $S^H_2$).}
    \label{fig:LSH}
\end{figure}
To recognize the super-period in the data, we use the Lomb-Scargle periodogram \citep{Gatspy2015}. The maximum power identified for planet c from Holczer data is around 600 days. In Figure~\protect{\ref{fig:LSH}}, we show the periodogram applied to the Kepler-59c data from Hoczer's catalogue, and applied to the same planet from the solutions $S^H_1$ and $S^H_2$. Superimposed in the plot are the analytic super-periods calculated before. 

%%%%%%%%%%%%%%%%%%%%%%%%%%%%%%%%%%%%%%%%%%%%%%%%%%%%%%%%%%%%%%%%%%%%%%%%%%%%%%%%%%%%%%%%%%%%%%%%%%%%%%%%%%%%%%%%%%%%%
\subsection{Kepler-59b and Kepler-59c TTVs}
\label{Results2}

\citet{Rowe2015} present TTVs for both planets, Kepler-59 b and c, obtained from quarters Q1 to Q12 of the Kepler mission. In this case Rowe identifies 71 transits for planet c and 110 transits for planet b. 

We apply the inversion method to the Rowe's data  and simultaneously fit the TTV signals of both planets. The {\tt MultiNest} sampling parameters are taken the same as in Section \ref{results1}. The result is a bimodal posterior for the planetary system with higher values of the evidence $\ln(Z)$ in comparison to the solutions fitting only one planet, eg,  $\ln(Z)_{S^R_1}=335.4$ and $\ln(Z)_{S^R_2}=337.8$. To compare the two solutions we apply the BIC criteria for each data set. The analysis shows $\Delta~BIC>10$, i.e., a strong evidence against the solutions obtained from Rowe's data. 

%can not apply the occam's razor because the data set is not the same in each case

Table \ref{tab:param59R} shows the results for both solutions. The first solution $S^R_1$ presents an inner planet in an orbit with a period of $P_b=11.8715_{-0.0005}^{+0.0005}$ days and a mass of $M_b=5_{-2}^{+4}~M_{\mathrm{\oplus}}$, while the outer planet has an orbit with $P_c=17.9742_{-0.0009}^{+0.0013}$ days and a mass of $M_c=4.6_{-2.0}^{+3.6}~M_{\mathrm{\oplus}}$. The eccentricities and mutual inclination of the planets indicate that they are in nearly circular and nearly co-planar orbits. This configuration generates the TTV signal shown in blue line in Figure~\protect{\ref{fig:O-C59Hc}. 

\begin{table*}
	\caption{Parameters estimated by solution 1 ($S^R_1$) and solution 2 ($S^R_2$) from \citet{Rowe2015}.
    The orbital parameters are the osculating astrocentric elements at epoch BJD~2\,455\,052. The upper block are the mean values and the error bars reported from the dynamical fit parameters, at the standard 68.34\% confidence level. The lower block reports the derived parameters.}
	\label{tab:param59R}
    \centering  
             \begin{tabular}{lcccc}
             \toprule
 &\multicolumn{2}{|c|}{$S^R_1$}  & \multicolumn{2}{|c|}{$S^R_2$}  \\ [0.5ex]
 & \bf{Kepler-59b}  &  \bf{Kepler-59c}  & \bf{Kepler-59b}  &  \bf{Kepler-59c} \\ [0.5ex] % inserts table
               \midrule
              Dynamical fit            &  &     \\ [0.5ex] % inserts table
                $M_p/M_\star\,(\times 10^{-3})$  & $0.11^{+0.08}_{-0.04}$ & $0.10^{+0.08}_{-0.04}$ & $0.07^{+0.01}_{-0.01}$ & $0.06^{+0.02}_{-0.02}$ \\
                \vspace{0.1cm}
                $P_p$~(d)  & $11.8715^{+0.0005}_{-0.0005}$ & $17.9742^{+0.0013}_{-0.0009}$ & $11.8714^{+0.0004}_{-0.0004}$ & $17.9737^{+0.0008}_{-0.0008}$ \\
                \vspace{0.1cm}
                $e_p$  & $0.05^{+0.09}_{-0.03}$ & $0.05^{+0.08}_{-0.03}$ & $0.09^{+0.09}_{-0.05}$ & $0.09^{+0.08}_{-0.05}$  \\
                \vspace{0.1cm}
                $b_p$  & $0.5^{+0.3}_{-0.3}$ & $0.5^{+0.3}_{-0.3}$ & $0.5^{+0.3}_{-0.3}$ & $0.5^{+0.3}_{-0.3}$ \\
                \vspace{0.1cm}
                $\varpi_p$~($^\circ$) & $309^{+148}_{-86}$ &  $21^{+57}_{-62}$ & $301^{+150}_{-69}$ &  $29^{+51}_{-89}$\\
                \vspace{0.1cm} 
                $\Omega_p$~($^\circ$) & $273^{+109}_{-106}$ & $270$  & $271^{+110}_{-112}$  & $270$  \\
                \vspace{0.1cm} 
                $\delta t$~(d) & $1.079^{+0.008}_{-0.008}$  & $0.004^{+0.004}_{-0.002}$ & $1.079^{+0.008}_{-0.008}$  & $0.004^{+0.005}_{-0.003}$ \\    
               \midrule        
             Derived parameters    &  &     \\ [0.5ex] % inserts table 
                \vspace{0.1cm}                
                $M_p$~($M_{\earth}$)  & $ 5^{+4}_{-2} $ & $4.6^{+3.6}_{-2.0}$  & $3.0^{+0.8}_{-0.8} $ & $2.6^{+0.9}_{-0.8} $  \\
                \vspace{0.1cm}                               
                $a_p$~(au) $0.112^{+0.002}_{-0.002}$ & $0.148^{+0.002}_{-0.002}$ & $0.112^{+0.002}_{-0.002} $ & $0.148^{+0.002}_{-0.002} $ \\     
                \vspace{0.1cm} 
                $i_p$~($^\circ$) & $88.38^{+1.07}_{-1.08}$ & $88.8^{+0.8}_{-0.8}$ & $88.4^{+1.1}_{-1.0}$ & $88.8^{+0.8}_{-0.8}$ \\   
                \vspace{0.1cm}                                 
                $I_p$~($^\circ$) & $1.66^{+1.07}_{-1.09}$ & $1.6^{+0.8}_{-0.8}$ & $1.6^{+1.2}_{-1.1}$ & $1.6^{+0.8}_{-0.8}$ \\   
                \vspace{0.1cm}                                 
                $I_{\mathrm{mut}}$~($^\circ$) & -- & $0.1^{+1.3}_{-0.1}$ & -- & $0.04^{+1.40}_{-0.04}$ \\   
                \vspace{0.1cm}                                 
                $R_p$~($R_{\earth}$)  & $1.5^{+0.1}_{-0.1}$ &  $2.2^{+0.1}_{-0.1}$ & $1.5^{+0.1}_{-0.1}$ &  $2.2^{+0.1}_{-0.1}$ \\
                \vspace{0.1cm}
                $\rho_p$~(g\,cm$^{-3}$)  &  $8^{+6}_{-4}$ & $2.4^{+2}_{-1.5}$ &  $4.9^{+1.3}_{-1.9}$ & $1.4^{+0.5}_{-0.6}$ \\
             \bottomrule               
               \end{tabular}                
\end{table*}

\begin{figure*}
    \centering
	\includegraphics[width=6cm,trim={3.2cm 0 3.5cm 0cm}]{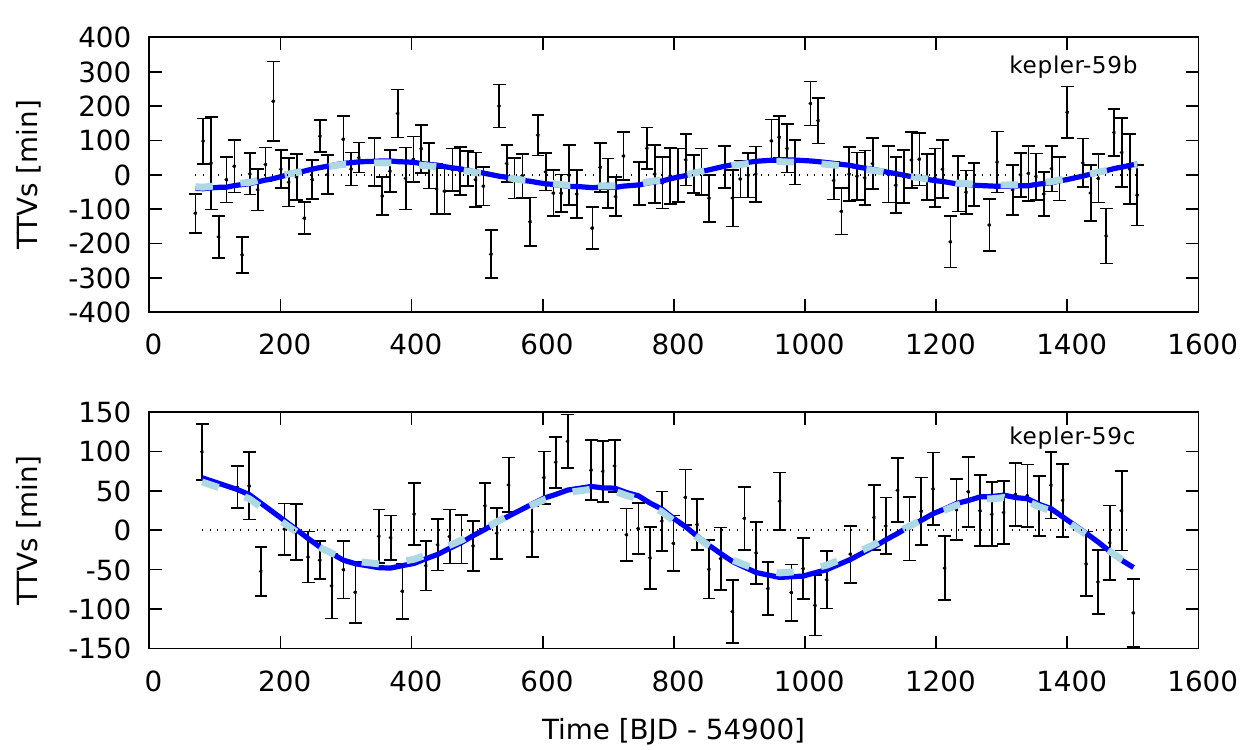}
	\caption {TTVs reported by \protect\citet{Rowe2015} as black dots with error bars. The best fit of solution $S^R_1$ is shown in blue, and the best-fit of $S^R_2$ is shown in cyan.}
    \label{fig:O-C59Hc}
\end{figure*}

The second solution, hereafter $S^R_2$, shows  planetary periods indistinguishable within 1-$\sigma$ from those of solution $S^R_1$ and a relation between them indicating a proximity to the 3:2 MMR. In this case the estimated masses are lower than in the first solution, $M_b=3.0^{+0.8}_{-0.8}~M_{\mathrm{\oplus}}$ for the inner planet and  $M_c=2.6^{+1.9}_{-0.8}~M_{\mathrm{\oplus}}$ for the outer planet. The eccentricities and inclinations do not show differences with respect to the first solution. The TTV signal produced from this solutions is shown as a cyan line in Figure~\protect{\ref{fig:O-C59}}.}

\subsubsection{TTVs periodicity }
\begin{figure}
    \centering
	\includegraphics[width=8.0cm]{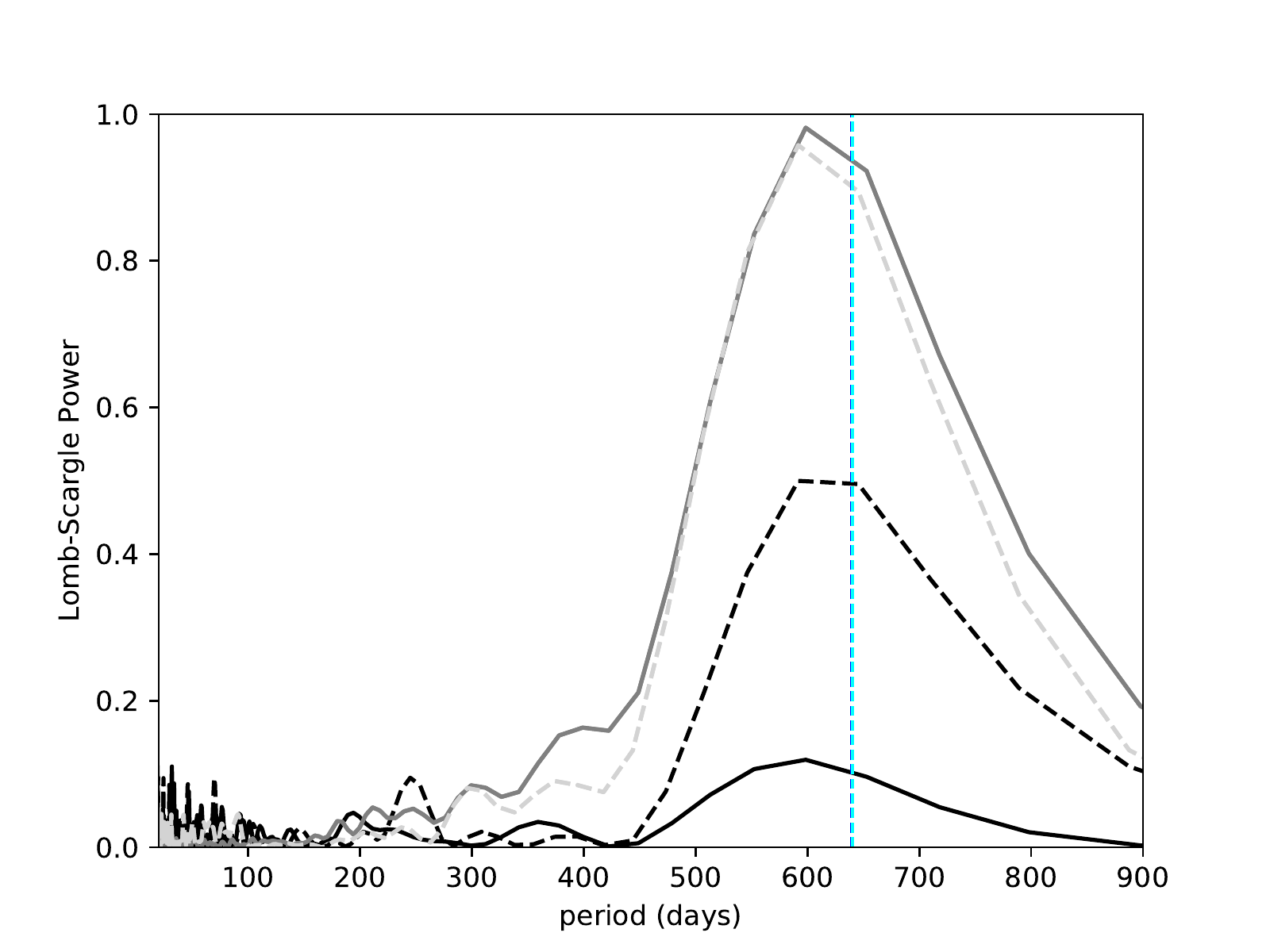}
	\caption {Power spectrum period of Kepler-59b (black full line), Kepler-59c (black dashed line), solution $S^R_1$ (grey full line for planet b and grey dashed line for planet c), and solution $S^R_2$ (light grey full line for planet b and light grey dashed line for planet c). Both solutions have similar periodograms for each planet. The dashed vertical lines are the analytic super-period estimates of both solutions (blue line for $S^R_1$ and cyan line for $S^R_2$).}
    \label{fig:LSR}
\end{figure}

We calculate the super-periods for the Rowe's data solutions, $S^R_1$ and $S^R_2$, following Eq. (\ref{eq:superperiod}):
\begin{equation}
    P_{S^R_1}=639.0~ \mathrm{d} \quad\ \text{ and } \quad\ P_{S^R_2}=640.3~ \mathrm{d}
\end{equation}
Then, we apply the Lomb-Scargle periodogram \citep{Gatspy2015} to the signals produced by the two solutions and to the data. The maximum power identified on Rowe's data is around 590 days for planet b and 610 days for planet c. In Figure~\protect{\ref{fig:LSR}}, we show the periodogram applied to Rowe's data as black lines, and that applied to $S^H_1$ and $S^H_2$ as grey lines. Superimposed are the super-periods calculated analytically. 

%%%%%%%%%%%%%%%%%%%%%%%%%%%%%%%%%%%%%%%%%%%%%%%%%%%%%%%%%%%%%%%%%%%%%%%%%%%%%%%%%%%%%%%%%%%%%%%%%%%%%%%%%%%%%%%
\subsection{Planetary densities}
As we found in Section~\protect{\ref{radius}} and taking into account the periods obtained here, planets b and c are near the {\tt Fulton gap}, just around the minimum occurring at $\sim$1.8~$R_{\mathrm{\oplus}}$ in the radius distribution \citep{Fulton2017, Martinez2019}.
Planets around this gap are thought to be either gas dwarfs consisting of rocky cores embedded in H2-rich gas envelopes, or water worlds containing significant amounts of H2O-dominated fluid/ice in addition to rock and gas. There are two possible mechanisms that can explain the presence of the high-density super-Earth on one side of the gap and the low-density sub-Neptunes on the other side: (1) evaporation due to high-energy stellar photons \citep{Owen2013,Owen2017,Wu2019}, and (2) the core-powered mass-loss \citep{Ginzburg2016,Ginzburg2018,Gupta2019}. The first process is more common during the early stages of the star formation, and may remove the light molecular weight envelopes of planets. In the second mechanism, the planet's internal luminosity produced from it's primordial formation energy drives the loss of its atmosphere.

\begin{figure*}
    \centering
	\includegraphics[width=6cm,trim={2.5cm 0cm 3.5cm 0cm}]{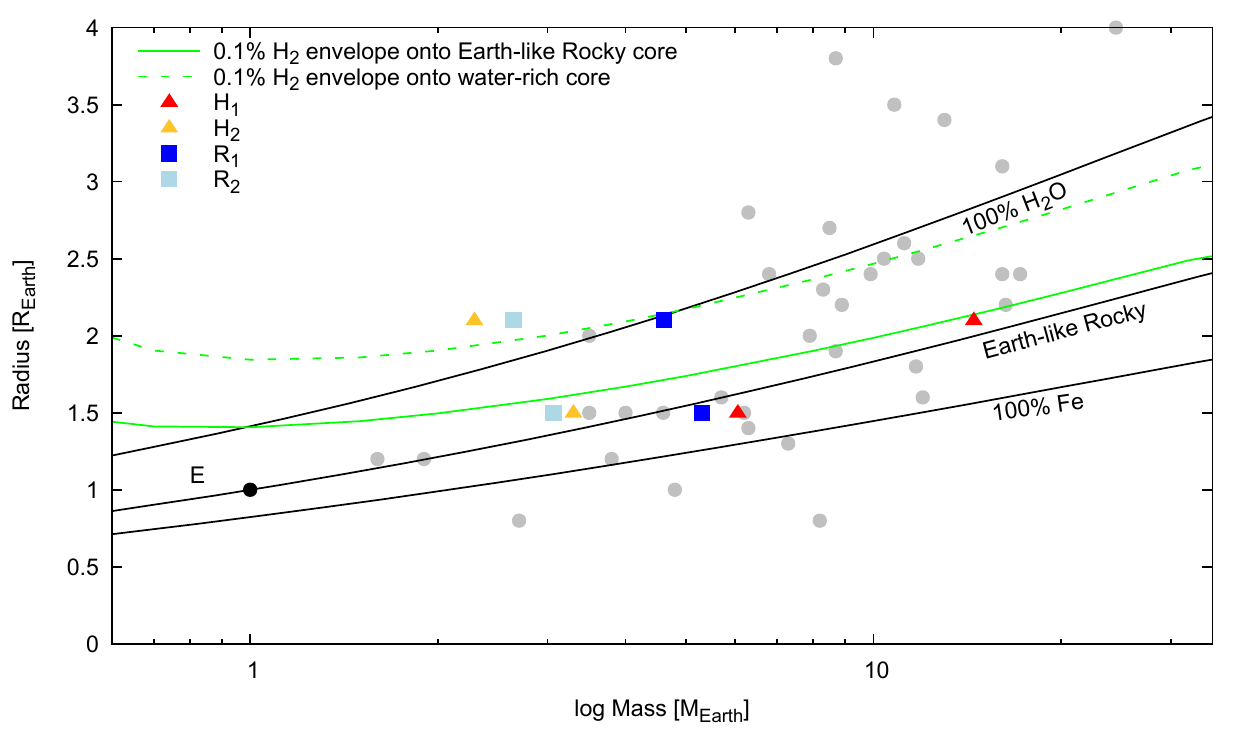}
	\caption {Mass-radius relationship for the two pairs of solutions found in this work. Solutions from Holczer's data are shown as triangles, $R_p\sim2.2~R_{\mathrm{\oplus}}$ for planet c, and $R_p\sim1.5~R_{\mathrm{\oplus}}$ for planet b: $S^H_1$ is shown in red and $S^H_2$ in orange. Solutions from Rowe's data are shown as squares, $R_p\sim2.2R_{\mathrm{\oplus}}$ for planet c, and $R_p\sim1.5R_{\mathrm{\oplus}}$ for planet b: $S^R_1$ is shown in blue and $S^R_2$ in cyan. Solid lines represents the theoretical two-layer mass-radius models from \citet{Zeng2019}, assuming single composition. For comparison, planets with well known masses and radii are shown as grey dots \citep{Hadden2017}, and Earth is shown as a black dot.
	}
    \label{fig:mr}
\end{figure*}

In order to calculate the densities of the Kepler-59 planets we consider the masses obtained from solutions $S^H_1$, $S^H_2$, $S^R_1$ and $S^R_2$, together with the planetary radii calculated in Section~\protect{\ref{radius}}. We present the inferred planetary radii and densities in Tables~\protect{\ref{tab:param59H}} and ~\protect{\ref{tab:param59R}}. The two solutions for each data set are represented in a diagram of planetary mass vs. planetary radius, shown in Figure~\protect{\ref{fig:mr}}. The diagram includes single composition lines taken from a two-layer mass-radius planetary model by \citet{Zeng2019}. The solutions from Holczer's data, $S^H$, are displayed as triangles, and those from Rowe's data, $S^R$, are displayed as squares. The inner planet (red triangle for $S^H_1$ and orange square for $S^H_2$) is located in the region where planetary composition is purely rocky. This fact is consistent with previous results that showed that the densities of planets with radii smaller than $\sim1.6~R_{\mathrm{\oplus}}$ are generally consistent with a purely rocky composition. On the other hand, most planets larger than $1.6~R_{\mathrm{\oplus}}$ have low weighted mean densities that are inconsistent with a rocky composition, and the decrease in density must be due to an increasing fraction of volatiles, with the secure presence of gaseous H/He  envelopes \citep{Seager2007, Rogers2015}. The solutions $S^H_2$, $S^R_1$ and $S^R_2$ are in agreement with these studies. The solution $S^H_1$ does not show the presence of any envelope around the planet. 

\citet{Chen2017} developed a new power law model for the mass-radius relation of exoplanets, considering three and four different types of objects: Terran, Neptunian, Jovian and Stellar worlds \footnote{Terran worlds may have oceans, ices, and/or atmospheres, but don't have a H/He envelope around them. These recall the inner planets in our Solar System. Neptunian worlds are dominated by a large atmosphere of hydrogen, helium, and other atoms/molecules that are easily boiled-off. They may have rocky interiors, but they obey a different mass/radius relationship than the Terran worlds. These planets recall Saturn, Uranus and Neptune.}. Their model uses the hierarchical Bayes theory to obtain the radius giving the mass, or vice-versa, using hyper-priors. 

To obtain more information to determine a unique solution, we apply the open code {\tt forecaster} that follows the model, and obtained that the planetary masses provided by solution $S^H_1$ produce the following radii: $R_b=2.77^{+0.98}_{-1.13}~R_{\mathrm{\oplus}}$ and $R_c=4.37^{+2.13}_{-2.38}~R_{\mathrm{\oplus}}$. For the masses provided by solution $S^H_2$, the planetary radii are: $R_b=1.70^{+0.44}_{-0.73}~R_{\mathrm{\oplus}}$ and $R_c=1.21^{+0.30}_{-0.69}~R_{\mathrm{\oplus}}$. We can see that, for each solution, the value of $R_b$ is fully compatible to the value inferred in our work, within the 1-$\sigma$ errors. For solution $S^H_1$, the value of $R_c$ is also compatible to our value within 1-$\sigma$, but for solution $S^H_2$ the values are only comparable within 2-$\sigma$ errors. Applying the same procedure to the masses provided by solutions $S^R_1$ and $S^R_2$, the code gives radii of $R_b=2.22^{+1.30}_{-1.11}~R_{\mathrm{\oplus}}$, $R_c=1.79^{+1.46}_{-0.61}~R_{\mathrm{\oplus}}$, $R_b=1.51^{+0.70}_{-0.30}~R_{\mathrm{\oplus}}$ and $R_c=1.45^{+0.55}_{-0.37}~R_{\mathrm{\oplus}}$, respectively. In this case, for each solution $R_b$ is compatible to our value within 1-$\sigma$, while $R_c$ is compatible within 2-$\sigma$.
Unfortunately, this analysis does not give any information on whether any solution $S^H_1$, $S^H_2$, $S^R_1$ or $S^R_2$ should be preferred over the others. 

\begin{figure*}
    \centering
	\includegraphics[width=6cm,trim={2.5cm 0cm 3.5cm 0cm}]{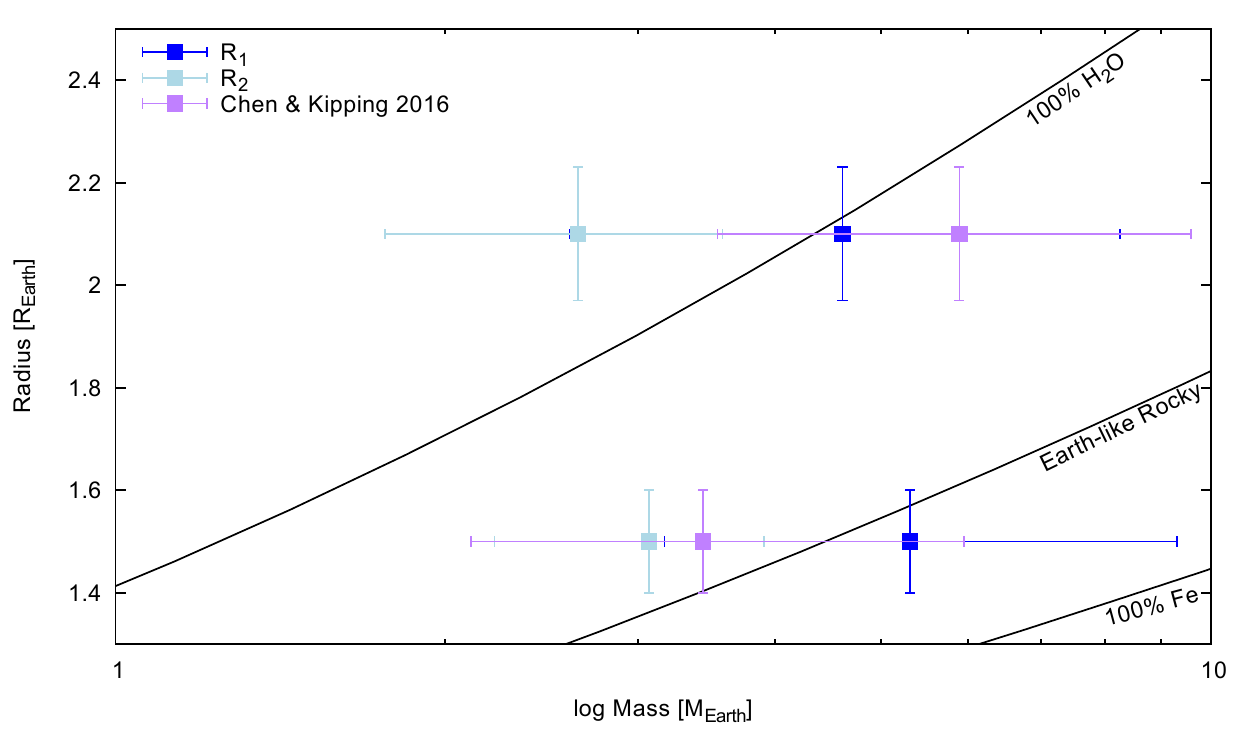}
	\caption {Mass-radius relationship for the solutions from Rowe's data (solution $S^R_1$ in blue and $S^R_2$ in cyan) and the empirical solution from \citet{Chen2017} model (in purple). Solid lines represents the theoretical two-layer mass-radius models from \citet{Zeng2019}, assuming single composition.
	}
    \label{fig:mr2}
\end{figure*}

\begin{figure*}
    \centering
	\includegraphics[width=6cm,trim={3.2cm 0 3.5cm 0cm}]{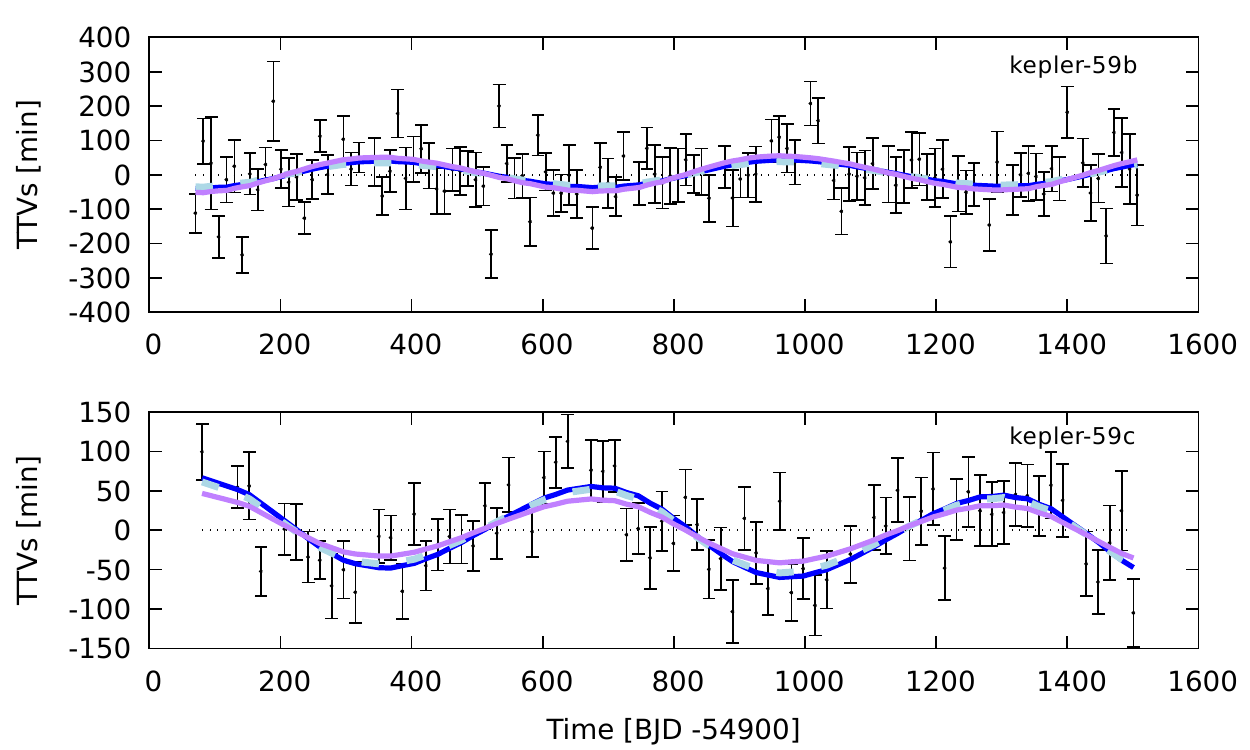}
	\caption {TTVs reported by \protect\citet{Rowe2015} (black dots with error bars) in comparison to the TTV signal obtained from \citet{Chen2017} model (in purple). The best fit of solution $S^R_1$ in blue and the best fit of solution $S^R_2$ in cyan. }
    \label{fig:O-C59-forecaster}
\end{figure*}

We take the planetary radii as input of the model and obtain a planetary mass of the inner planet of $M_b=3.43^{+2.51}_{-1.32}~M_{\mathrm{\oplus}}$ and for the outer planet a mass of $M_c=2.64^{+0.94}_{-0.88}~M_{\mathrm{\oplus}}$. In Figure \ref{fig:mr2} we add this empirical solution (purple dots) to the mass-radius correlation compared to the solutions $S^R_1$ and $S^R_2$ (blue and cyan dots, respectively).
The comparison confirms that Kepler-59 is a system with an inner super-Earth planet and an outer mini-Neptune. The solutions $S^R_1$ and $S^R_2$ show that the outer planet seems to be less massive than the inner one. But there is a difference between the slope that the radius-to-mass relation of this system maintains for these solutions to the result obtained from Chen's model, showing an opposite behaviour. 
This difference can also be observed when the mid-transit times are generated. Figure \ref{fig:O-C59-forecaster} shows in purple lines the signals calculated for $S^R_1$ and $S^R_2$ parameters and for the masses obtained with {\tt forecaster}. The amplitude of Kepler-59c TTV signal is lower than that obtained from this work planetary masses (blue and cyan lines). For planet Kepler-59b there is an increase in the amplitude signal. This is also visible when comparing the $\chi^2_\mu$, being $\mu_b=109$ for planet b and $\mu_c=58$ for planet c. The fit with Chen's model gives two times the value of the solutions obtained in this work ($\chi^2_\mu\sim1.6$ for both planets in both solutions).
 
%%%%%%%%%%%%%%%%%%%%%%%%%%%%%%%%%%%%%%%%%%%%%%%%%%%%%%%%%%%%%%%%%%%%%%%%%%%%%%%%%%%%%%%%%%%%%%%%%%%%%%%%%%%%%%%
\subsection{Dynamical stability}
\label{dynamics}
We performed an analysis of the stability of the Kepler-59 system by means of long term N-body simulations and the construction of stability maps. The simulations have been carried out using the module \texttt{Helio}, available as part of the \texttt{Swifter} integration package (\citealp{Levison1994}; Kaufmann \& Levison, \url{http://www.boulder.swri.edu/swifter/}). The simulations involve a system of two planets around the star, and use  the nominal values of the orbital elements and masses of the four solutions found here as initial conditions. The total time span was $10^8$~yr, with a time step of 0.1 days. The evolution of the orbital elements of the planets indicate that the system is stable over the whole time span, no matters which solution is chosen.

We have also constructed dynamical maps to assess the stability of the system in the $m_b$, $m_c$ parameters space. We followed the evolution of a uniform grid of $64\times64$ initial conditions, covering the mass ranges provided by the errors of solution $S^H_1$, which shows the largest range of masses among the four solutions. For these simulations, we apply a modified version of the \texttt{Swifter Helio} module, translated into CUDA-C to run on a GPU
architecture (Costa de Souza et al., in preparation). This code computes several stability/chaos indicators along the simulations, which are then translated into a color scale to construct the maps. We found, once more, that the system is stable along the whole range of tested masses.

%%%%%%%%%%%%%%%%%%%%%%%%%%%%%%%%%%%%%%%%%%%%%%%%%%%%%%%%%%%%%%%%%%%%%%%%%%%%%%%%%%%%%%%%%%%%%%%%%%%%%%%%%%%%%%%
\section{Conclusions}
\label{discu}
In this work, we have characterized the masses and radii of the star and planets in the Kepler-59 system, as well as their orbital parameters. Our methodology was based on the inversion of TTVs signals applying a Bayesian inference tool to determine model parameters. 

Recently works have shown that when com-pared to radial velocity analyses TTVs let us know with lesser precision, but not unambiguously, the mass and eccentricity of the planets. All this is true in the case of having a high quality of TTV (S/N, quantity of mid-transit times observed) and in favourable dynamical configuration of the systems (Saad-Olivera et al. 2017; Saad-Olivera et al. 2018). Here, we showed once again the power of TTVs for the determination of planetary masses. Despite this case needs a better transit data reduction in order to gain more S/N in the mid-transit times of both transiting planets.

We have considered two different sets of TTV data: one set provided by \citet{Holczer2016}, with TTVs only for Kepler-59c, and the other set provided by \citet{Rowe2015}, with TTVs for both Kepler-59b and c. For each of the data sets, we found two possible solutions having the same probability according to their Bayesian evidences. All the four solutions appear to be indistinguishable within their 2-$\sigma$ uncertainties, but the solutions from Rowe's data display larger values of the evidences, due to the use of transit information from the two planets.

Our results point out to a system with a super-Earth in an inner orbit and a mini-Neptune in an outer orbit. The planets lie in almost co-planar, almost circular orbits ($e<0.1$). Their periods ratio put them close to the outer border of the mutual 3:2 mean motion resonance, but not inside this resonance. Stability analysis indicate that this configuration is stable over the long term.

The derived densities of the planets and periods imply that the planets are around the radius gap known for planets <$4R_{\mathrm{\oplus}}$. This turns Kepler-59 system to be a great laboratory to study the conditions that creates the Fulton gap. Following the photo-evaporation model it can be said that the innermost planet probably has a rocky core that would have lost its envelope of H/He during its formation. On the other hand, the outer planet may still retain an envelope of light elements and volatiles. Further studies are needed to determine the nature of these planets.

\section*{Acknowledgements}
The authors thanks the anonymous referee for the constructive comments.
The simulations have been performed at the SDumont cluster
of the Brazilian System of High Performance Computing (SINAPAD). This work has been supported by the Brazilian National Council of Research (CNPq), by the Brazilian Federal Agency for Support and Assessment of Postgraduate Education (CAPES), and by NASA's XRP Program. This work has made use of data from the European Space Agency (ESA) mission Gaia (https://www.cosmos.esa.int/gaia), processed by the Gaia Data Processing and Analysis Consortium (DPAC, https://www.cosmos.esa.int/web/gaia/dpac/consortium). Funding for the DPAC has been provided by national institutions, in particular the institutions participating in the Gaia Multilateral Agreement.
 
%%%%%%%%%%%%%%%%%%%% REFERENCES %%%%%%%%%%%%%%%%%%

\bibliographystyle{mnras}
\bibliography{kep59} 

%%%%%%%%%%%%%%%%%%%%%%%%%%%%%%%%%%%%%%%%%%%%%%%%%%

\bsp	% typesetting comment
\label{lastpage}
\end{document}